# An efficient tight-binding mode-space NEGF model enabling up to million atoms III-V nanowire MOSFETs and TFETs simulations.

A. Afzalian, T. Vasen, P. Ramvall, M. Passlack
TSMC, Leuven, Belgium,

*Abstract* — **We report the capability to simulate in a quantum mechanical tight-binding (TB) atomistic fashion NW devices featuring several hundred to millions of atoms and diameter up to 18 nm. Such simulations go far beyond what is typically affordable with today's supercomputers using a traditional real space (RS) TB Hamiltonian technique. We have employed an innovative TB mode space (MS) technique instead and demonstrate large speedup (up to 10,000×) while keeping good accuracy (error < 1%) compared to the RS NEGF method. Such technique and capability open new avenues to explore and understand the physics of nanoscale and mesoscopic devices dominated by quantum effects.**

**In particular, our method addresses in an unprecedented way the technological relevant case of band-to-band tunneling (BTBT) in III-V nanowire MOSFETs and broken gap heterojunction tunnel-FETs (TFETs). We demonstrate an accurate match of simulated BTBT currents to experimental measurements in a [111] InAs NW having a 12 nm diameter and a 300 nm long channel. We apply the predictivity of our TB MS simulations and report an in-depth atomistic study of the scaling potential of III-V GAA nanowire heterojunction n and pTFETs quantifying the benefits of this technology for low-power, low-voltage CMOS application. At $V_{DD}$ = 0.3 V and $I_{OFF}$ = 50 pA/μm, the on-current ($I_{on}$) and energy-delay product (ETP) gain over a Si NW GAA MOSFET are 58× and 56× respectively.**

Decades of technological progress in micro and nano fabrication techniques have enabled the fabrication of solid-state devices with ever smaller dimensions. This scaling has been dictated by Moore's law,[1] which has been the main driving force for the CMOS industry for over 4 decades and has enabled rich scientific breakthrough and large societal benefits. Typical device dimensions, today, range from a few nm to tens of nm, making the pursuing of the dimension scaling increasingly difficult.[2] As a consequence, the semiconductor devices are evolving from planar MOS structure to 3D multigated transistors such as FinFETs or nanowires (NW).[2,3,4] New channel materials, such as Ge,[4] III-V[5] or 2D materials[6] as well as new computing and device concepts are being sought after.[2] Amongst other power consumption and heat dissipation has become a key concern in modern CMOS technologies due to the ever increasing number of transistors per unit area and new energy efficient switches are sought after as replacement to the MOSFET concept.[7,8]

In a MOSFET the lower limit for the subthreshold slope, or the inverse slope of the drain current – gate voltage characteristic $I_D(V_G)$ of a transistor, is $ln(10) \times kT/q$, i.e. at least 59.6 mV of gate voltage variation is required to change the current by a decade at 300K, sets a practical limit to reductions of supply voltage and power consumption of a circuit.[7,8] Achieving steeper subthreshold slope transistors has become a key concern for further CMOS downscaling and different concepts to overcome this limitation have been proposed. This includes devices using a multiphysic effect or feedback mechanism triggered by $V_G$ such as electromechanical motion in the Suspended-Gate (SG)-MOSFET[9] or replacing the standard gate insulator with a ferroelectric insulator of the right thickness in the ferroelectric FET.[10,11] Other concepts attempt to use impact ionization,[12] or energy filtering with tunnel barriers and resonant tunneling in the RTFET,[13] by means of a superlattice to create minibands in the Superlattice FET,[14] or by using band-to-band tunneling through gate modulation of reverse-biased PN junctions, in a Tunnel-FET (TFET).[15,8] The TFET is probably the most mature steep slope device technology today.[8] In particular, III-V gate-all-around (GAA) nanowire (NW) heterojunction TFETs (HTFET) can potentially replace conventional Si MOSFETs as a low power (LP) sub-10 nm technology option.[16,17,18] Despite the progress in the field,[16] no experimental verification of sub-10 nm diameter ($d$) GAA NW III-V HTFET performance has been reported.

In front of the large panel of possible technological options, predictive simulations that could accurately assess the performance and viability of these technology options at scaled dimension are highly desirable. Semi-classical transport models,[19] on which traditional simulation approaches are based, allow for fast computation. They however rely on macroscopic parameters (such as mobility) that require calibration and do not naturally include quantum effects which are strongly present in today's nanoscale solid-state devices. In order to explore the properties of nanoscale devices, or layers of innovative materials and their interface with semiconductors, one needs device-level models and simulation tools that can directly infer device properties from their atomic structure. Among the adequate microscopic device simulation techniques, the NEGF (Non-Equilibrium Green's Function) method has gained popularity and shown a real possibility to capture the essential physics at these scales.[20,21,22] In the NEGF approach, the input microscopic parameters are introduced via the Hamiltonian ($H$). In particular, full-band atomistic tight binding (TB) Hamiltonians only require well tested and transferable orbital bases for the atoms[23,24,25] as input and can be considered as the golden reference.

The computational cost of these atomistic simulations is however heavy requiring massive parallel computational resources and typically limited to devices of very small dimensions. Specifically, in devices where the cross-sectional dimensions are too small to be supposed infinite so that Bloch periodic condition simplification cannot be used, as it is typically the case for nanowires, nanotubes, or nanoribbons, the full (3D) device atomistic structure need to be simulated, leading to a simulation time and memory requirements that increase exponentially with the cross-section dimensions (Fig. 1). In particular, for simulating band-to-band tunneling (BTBT) in III-V NW TFETs with technology relevant dimensions, i.e. $d$ in the 4 - 10 nm range, a larger atomic basis including spin orbit coupling (SO), such as $sp^3s^*SO$ (10 orbitals/atom) is needed to ensure correct band gap and valence band (VB) curvatures in III-V materials[26,27] (Fig. 2b). Simulation time increases with a power-3 law with respect to atomic basis size and a power-6 law with NW diameter. As a result, small diameter NW ($d \sim 2$ nm)[28] and/or a less accurate smaller non-SO III-V basis including $sp^3s^*$ (5 orbitals/atom),[26] $ss^*$ (2 orbitals/atom)[17] or $p_z$ carbon (1 orbital/atom)[29] have been used to predict performances of III-V NW devices. BTBT current is however extremely sensitive to band gap and band edge dispersion characteristics (Fig. 2b).

We report here the capability to simulate in a quantum mechanical tight-binding (TB) atomistic fashion NW devices featuring several hundred to millions of atoms and diameter up to 18 nm (Fig. 1). Such simulations go far beyond what is typically affordable with today's supercomputers using a traditional real space (RS) TB Hamiltonian technique. We have employed an innovative TB mode space (MS) technique instead and demonstrate large speedup (up to 10,000×) while keeping good accuracy (error < 1%) compared to the RS NEGF method (Fig. 2). Such technique and capability open new avenues to explore and understand the physics of nanoscale and mesoscopic devices dominated by quantum effects.

Using an incomplete basis, the MS approach has been widely used to significantly speed up effective mass (EM) NEGF simulations while keeping accuracy in a reduced energy window.[30-35] A direct transcription of the MS technique to TB models typically fails as the mode space band model is plagued with unphysical modes (UM) (Fig. 3 and Fig. 1 in SI).[24,36] A solution to this problem was proposed and its applicability to a few Si[24] and III-V homo- and heterojunction NW device cases[36,18] demonstrated. For each different NW cross-section slab (e.g. different material, diameter, channel orientation) a basis has to be derived, cleaned (Fig. 4, and Fig. 2 and 3 in SI) and stored in a database prior to the simulation. Such cleaning procedure is not straightforward and typically unsuccessful without a thorough understanding and control over the entire procedure. For this study, about 50 different bases were derived and speedup factors of up to 10,000× were achieved, which are by far the largest ever reported, demonstrating the wide applicability and benefits of the technique. In the 1st part of the result section, we will detail the critical steps needed to ensure an accurate and compact basis. We will also benchmark the TBMS technique in terms of speed, memory and accuracy against the RS NEGF method for III-V MOSFET and HTFET with diameter ranging from 4 to 18 nm. The latter being by far the largest NW atomistically simulated reported in the literature to the best of our knowledge. Comparison to experimental III-V nanowire MOSFET with a good match between measured and modeled BTBT current will also be shown.

In the second part of the result section, we apply the technique to investigate the physics and performances of III-V heterojunction broken gap TFETs. Our method was implemented in the NEMO5 atomistic simulation tool.[25] It addresses in an unprecedented way the technological relevant case of band-to-band tunneling (BTBT), allowing for a wide range and number of TB atomistic NEGF simulations of III-V NW featuring several 100,000 to millions of atoms. We report here the first in-depth atomistic optimization study of III-V GAA NW HTFET from a scaling perspective with $d$ in the 4 - 10 nm range, $L$ in the 10 to 25 nm range and crystal orientation dependence. Impact of material choice and architecture is also considered by adding an InGaAs heterojunction. Our study highlights and quantifies the potential benefit of such technology. In particular an optimized $L$ = 20 nm [111] oriented NW GAA TFET design features $I_{ON}$ and energy delay product (ETP) performance gain over a Si NW MOSFET of 58× and 56× respectively.

**RESULTS:**

*A. Tight-Binding Mode Space Model*

The general principle of a low rank approximation NEGF method such as MS is to switch from a full representation space of size $N$, e.g. a real space Hamiltonian representation, to a space where a smaller but accurate subspace of size $n<<N$ can be found to solve a given problem, that is a representation space where one can distinguish between useful and non-useful information. As an example in RS the entries of the atomistic tight binding $H$ matrix represent atoms and their atomic orbitals and $N$ is related to the number of atoms × the number of atomic orbitals per atom. All the atoms are needed to represent a given device physics and reducing the number of orbitals results in reducing the physical accuracy of the model (Fig. 2b). By switching to the eigenmodes representation space in the mode space method however, one can choose a subset of modes $\Psi_i$ ($n<<N$) whose eigenvalues are closes to the band edges of the nanowire slab of interest and neglect higher energy modes in the Hamiltonian. Such a reduced model is both efficient and accurate (in a reduced energy window of interest), as further away energy modes do not participate to the electronic transport. The energy window of interest depends on the problem to solve. It can be described without loss of generality by $E_L$ and $E_H$, the lower and higher energy bound on which the model is to be accurate.

To switch from the original real space of size $N$ to the reduced mode space of size $n$, a unitary transformation basis composed of the $n$ chosen orthonormal basis eigenvectors { $\Psi_i$ } in the $N$-dimensional basis need to be constructed. Choosing $n$ representative basis vectors $\Psi_i$, a direct application of the MS technique to TB models usually fails. When constructing the MS transformed model, the mode space observables are plagued with unphysical modes (UM), even if the chosen basis reproduces well the physical modes of interest. This appears clearly in the band structure of such a MS model (Fig. 3 and Fig. 1 of SI). Using an optimization procedure, it is possible to enlarge the MS basis by adding new basis vectors specially "tailored" to remove the unphysical branches in the MS model.[24]

The complexity of cleaning a basis increases with the number of unphysical mode to eliminate. The latter typically increases with the number of desired physical subband modes, especially when these subbands are close in energy from each other. This is typically the case in the valence band (VB) of group III-V nanowire band structure for instance (Fig. 3 and Fig. 1 of SI). As a consequence at small diameters only a few unphysical modes are present after the initial sampling and these can be cleaned within a few optimization steps, while for larger diameters hundreds of unphysical modes and optimization steps will be required (Fig. 3 and 4). At same diameter a basis that includes only conduction subbands modes features much less UM than a basis including VB modes, while a basis that can reproduce both VB and conduction band (CB) modes will have the most. For the same reason the complexity and the number of UM to clean increases when increasing the number of orbitals per atom of the TB model (see Fig. 3).

In order to model band to band tunneling, it is however desired to derive bases that accurately reproduce CB and VB states and include a large number of orbitals per atom. As a result, in the bases we consider here, the number of unphysical modes to eliminate is very high (typically several hundreds) compared to the Si case reported in[24] which has only a few UM. Nevertheless, we demonstrate here the possibility to clean III-V MS basis from the UM (Fig. 4, and SI Fig. 2, 3 and 4) and achieve a significant speed up ratio (Fig. 1 and 2). In order to clean the basis, we shall use the functional minimization procedure proposed by Mil'nikov and coworkers[24]. We reproduce the main equations in the methods section for sake of completeness. We found that using this functional method typically fails, but using a few and simple to implement prescriptions. We will now revisit the optimization method, highlighting the critical steps to obtain accurate and efficient bases.

*1) Initial sampling:*

The first step that consists in creating an initial MS basis by sampling the NW unit structure or slab eigenmodes at selected energy $E$ - wave vector $k$ points is a crucial step upon which relies the successful derivation of an optimized basis. It will strongly determine the size and accuracy of the final basis. It is important to sample the physically relevant modes for the problem to be simulated, that is to determine the energy window that is needed for the simulation. Any modes out of the energy window of interest should not be sampled as they will typically increase the size of the final basis and increase the number of UM, hence the number of optimization steps and the complexity needed to clean the basis.

Successful initial samplings are illustrated in Fig. 3 and SI Fig. 1 for various InAs and GaSb NW slabs in a $sp^3s^*$ basis including Spin Orbit coupling. Concerning the density of wave vector points to be sampled, we typically choose a dense $k$ mesh for the modes of interest as depicted and eliminate any possible linear combination (eigenvectors with small $\Lambda_i$ values, see eq. (2) in the method section). We also note that it turns out to be important to sample the full Brillouin zone (BZ) and not just half of it despite the symmetry. This to avoid accuracy loss in the non-sampled half of the model, as showing up in the BS of the final optimized half BZ-sampled MS model (SI Fig. 5). In practice we typically alternate $k$ sampled values between each half BZ to get a more accurate result over the full BZ.

*2) Basis Optimization :*

The initial MS basis is not usable, as the MS transformed model is plagued with UM. A possible solution is to enlarge the MS basis by adding new basis vectors $\phi$ specially optimized to remove the unphysical branches in the MS model.[24] In practice, we add 1 basis vector at a time and only one or a few UM will be removed at each step. The technique need to be iterated up to total cleaning of the energy window of interest.

The new basis vectors are chosen by minimizing a functional $\Delta F$ (eq. (5) of the method section). $\Delta F$ measures the changes in the unphysical part of the band structure at $n_q$ chosen wave numbers $q_i$ brought about by the addition of 1 particular basis vector $\phi$. Selecting the optimization points is probably the most crucial point of the optimization procedure. In case with few UM, it has been shown that using a few $q_i$, e.g. $n_q=3$ and $q_i=0, \pi/2, \pi$, works.[24] In more complex cases, like the ones relevant for this study, however, 20 or more $q_i$ optimization points on the full BZ (e.g. from $-\pi/$ to $\pi$, see SI Fig. 6) are required and not properly selecting these points is the main reason for cleaning failure. As also shown on SI Fig. 6a, optimizing only on half the BZ (e.g. for positive $k$ modes) leads to residual UMs in the other half. Such a basis will provides correct transport properties for positive $k$ values only. Unless stated otherwise, all the bases derived here were sampled and optimized on the full BZ. Only then can we ensure full symmetry and accuracy for positive and negative $k$ values.

It is important to remind that the optimizer only "sees" (and removes) the UM located in the energy windows at the selected $q_i$ points. When adding a new mode to the basis, new UM can also be generated. Again the optimizer weights that impact at the $q_i$ points only. As a consequence, if one does not use enough $q_i$ points it is typical to see the optimizer moving one or a few "badly behaved" modes in between its optimization points, where they will never be removed. We found therefore that using a dense $q_i$ mesh ($\sim$ 20 regularly spaced points on the full BZ) in a first phase is best to remove most of the modes. It typically leads to the most compact bases, i.e. needing the fewer number of additional basis vectors, for a given choice of initial sampling. A second phase is often needed, especially for the largest bases. After the 1st phase (that typically last several tens to hundred of optimization steps) a few unphysical modes may remain. These UM typically feature a very narrow extend in $k$-space and dwell between the $q_i$ points so that the optimisation procedure is practically blind to them. By adjusting the $q_i$ mesh to increase the visibility of these modes, however, these UM will be swiftly removed.

Finally, we note that it is always possible to achieve a larger energy range from a previously optimized basis by performing an new initial sampling (e.g. adding new basis vectors at higher energy) to the previously optimized basis and perform a new optimization. However this is hardly necessary and we found that by properly choosing our initial sampling and optimization points, we can clean even very large bases with a single initial sampling.

*3) MS NEGF Performance and Accuracy Benchmark:*

We now assess the performance and accuracy of the MS method in the case of InAs NW MOS transistors and for InAs/GaSb Heterojunction TFETs in the ballistic case. We note that in case of electron-phonon scattering the overall simulation time increases both in RS and MS due to the fact that additional self-consistent Born iterations are needed to compute the scattering self-energies.[33,34] Efficient Form Factor methods have been developed for MS and benefits of the method in terms of speed up and memory reduction are therefore expected to be similar.[34]

*a) Accuracy*

Using the procedure described in the previous sections, MS bases for InAs and GaSb NW slabs with a square cross-section of size $d$ ranging from 4 to 18 nm were optimized from an initial $sp^3s^*$ SO TB basis and slab Hamiltonian sizes in the MS were a few % from their original size. Fig. 4 and SI Fig., 2, 3 and 4 compare the bandstructure of the optimized MS models to those of the original models for a few representative cases, showing a clean bandstructure in an energy range from a few 100 meV below the top of the VB and the bottom of the CB.

In order to fully validate and assess the accuracy of the mode space models, we simulate and compare $I_D(V_G)$ characteristics of [100] $d$ = 5.5 nm 15 nm long InAs MOSFET (Fig. 5a) and InAs-GaSb PNIN n-TFET (Fig. 2b) computed from the original (RS) TB model and from optimized InAs and GaSb MS bases (whose bandstructures are shown on Fig. 4c and SI Fig. 4a respectively). I-V curves obtained from the MS NEGF models match well (the accuracy is better than 1%) those obtained in the full model. Using the MS models, speed up factors larger than 150 were achieved, when compared to the initial TB models, for both the MOSFET and TFET cases (Fig. 5a and 2b). For the TFET case, the $I_D(V_G)$ from a RS model neglecting SO is also shown. Neglecting the SO coupling allows to reduce the RS Hamiltonian size by a factor 2, hence achieve a speed up factor of 8× compared to the RS SO model. It however removes an important part of the physics and the band structure is altered. In that case, an error larger than 50% is observed in the $I_D(V_G)$ characteristics (Fig. 2b). It is possible to alter the $sp3s*$ atomic parameters to compensate for the loss of physical accuracy and better fit the bandstructure of the SO model for a given diameter of interest ($d = 5.5$ nm in our case), hence improve the accuracy of $I_D(V_G)$ characteristics. Such parameter set is however not transferable to another diameter, while the speed up factor is comparably small compared to the MS speed up.

We also compared [111] $d = 12$ nm InAs NW nMOSFET experimental[37] vs. simulated MS ballistic data on Fig. 5b. A cross-section TEM image of a fabricated NW can be seen on Fig. 5c. More details about the fabricated NW including the fabrication process, process flow, SEM and TEM image can be found in the method section and on SI Fig. 7. Good agreement of the BTBT current and the subthreshold characteristics is observed assuming a diameter of $d = 12$nm at a drain Voltage $V_D = 0.5$V. We also observed as expected both in the simulation and the measurement the suppression of the BTBT current at $V_D = 0.3$V. This is easily explained by the fact that at a drain voltage bias below the bandgap of the NW in eV (which is about 0.45 eV for the $d = 12$ nm NW), the tunneling path between channel VB to drain CB is closed. The higher simulated current compared to the measured one in the on-state is easily explained by the ballistic hypothesis used in the simulation. From the comparison between simulation and experiments, we also extract a typical ballistic ratio of $r = 75\%$ for the 300 nm long InAs device.

*b) Performance:*

In the NEGF method, the simulation time is typically dominated by the inversion operations required to calculate the Green's functions $G^{20-22}$. The operation scales like $N_{tot}^3$, the cube of the Hamiltonian matrix size. $N_{tot}$ is in fact the total number of atoms in the device × the number of orbitals/atoms. Typically however, physical systems feature a limited interaction range and the system matrices are sparse. This sparcity is exploited in the recursive Green's function (RGF) algorithm[38,39] to considerably speed up the operation. In that case the scaling is typically cubic to the number of atoms in the cross-section, while rather linear in the number of atoms along the length direction. More details can be found in the method section (NEGF method and speed up consideration), including sparcity consideration and a derivation of the theoretical maximum speed up, $s_{max}$ (eq. (9)).

Fig. 1b, 1c and 2a show the typical simulation time /IV, peak memory usage/ core and speed up achieved on a 400 core cluster for III-V InAs MOSFET and InAs/GaSb TFETs devices with a diameter $d$ ranging between 4 to 18 nm and a total length $L_T = 100$ nm. The MS benefit strongly increases with $d$ as $r$ typically improves. The number of atoms increases as $d^2$, leading to a simulation time $t_{RS}$ increasing as $d^6$, while the memory required increases as $d^4$. The number of modes needed to accurately simulate a NW on the other hand increases slowly with $d$ so that $r$ improves.

Using the MS models, reduction ratio ranging from 8% (at $d = 4$ nm) down to 1% (at $d = 18$ nm) were observed and speed up factors $s$ ranging from 40 to 10,000, as well as memory reduction ranging from 4 to 100×, were achieved, when compared to the initial TB models, for both the MOSFET and TFET cases, using a cluster having 400 cores (Fig. 2a). This allows us to simulate NW with a diameter significantly larger than 10 nm and featuring several 100,000 and up to 1,000,000 of atoms with an atomistic 10 orbital/atom model, while this to the best of our knowledge it has never been reported before even on the world largest supercomputers.

Finally, we note that observed $s$ in our implementation are typically in the range of 40 to 60% of $s_{max}$ as some operations do not scale the same way as the Green's function inversion. In addition, in MS, there is typically an overhead linked to the down conversion from RS to MS of the device Hamiltonian including the latest electrostatic potential and the up conversion of the density to feed back the density to Poisson's equation for each NEGF – Poisson self-consistent loop.

*B. TB MS NEGF simulation of III-V Broken Gap NW TFETs*

We next apply our innovative TBMS NEGF technique to assess the physics and performance of III-V NW GAA broken gap HTFET. Fig. 6 shows the evolution of on-current ($I_{ON}$) vs. gate length ($L$) for optimized InAs/GaSb heterojunction n- and pTFETs and Si n MOSFETs when scaling $L$ from 25 to 10 nm. We focused here on fundamental performance and ideal devices were simulated. For the TFETs, we used ballistic simulations which have been shown to be a good approximation[17]. Scattering of electrons with acoustic and optical phonons was considered within the Self-Consistent Born approximation for the nMOSFETs using our in-house NEGF simulator *NANOcore*[34]. For each $L$ a full optimization, including diameter, channel orientation, doping, and drain underlap length dependency, was performed. Doping pockets were used to boost the performance of the TFETs.[40,41] More details on the device structure and impact of pockets can be found in the SI (text and Fig. 8, 9 and 10 of the SI).

As can be seen from Fig. 6, TFET performance degrades quite significantly and faster than that of Si MOSFETs when scaling $L$ below 20 nm. The main reasons for the stronger degradation in the TFET cases are identified as follows. 1) More pronounced short channel effects, in particular a "source-to-drain" tunneling (SDT) effect. As $L$ is reduced below 20 nm, an increasing part of the BTBT current is able to tunnel through the channel barrier to the drain in the off-state. In turn, this degrades the maximum achievable on-current. 2) A NW diameter of about 5.5 nm, which is optimal for controlling the TFET short channel effects at $L = 20$ nm, also yields optimally balanced InAs/GaSb material properties for the TFET application (Fig. 7 and SI Fig. 11). Due to quantum confinement, III-V EM, DOS, and bandgap (BG) increase. This is typically beneficial for improving subthreshold swing (SS) and reducing the ambipolar leakage current, but affects the tunneling probability in the on-state, $T_{ON}$ and reduces $I_{ON}$ when further scaling down $d$, especially for the pTFET case as discussed below.

Next, we focus on the design of n and p GAA NW TFETs for $L = 12$ and $L = 20$ nm with target values of off–current and supply voltage for LP applications ($I_{OFF} = 50$ pA /μm and $V_{DD} = 0.3$V) to illustrate design trade-offs and implications. Fig. 8 shows the impact of channel orientation for n- and pTFETs at both gate lengths for NW $d$ of 5.5 nm. With confinement, bandstructure anisotropy is enhanced (SI Fig. 3 and 4) and channel orientation has a strong impact on device performance. Both for n and p type devices, the [100] channel orientation features the largest effective band gap and relatively large effective masses, which reduce SDT and lead to the steepest SS in most cases, but also reduces $T_{ON}$. As a consequence at $L = 20$ nm, the best performances for both n- and pTFETs are achieved for [111] orientation that features the smallest effective bandgap and a low GaSb hole effective mass. The [110] orientation features the smallest electron and hole effective masses. Consequently it presents improved performance (comparable to the [111] case) for the nTFET, but worst pTFET performance due to SDT degradation. At $L = 12$ nm, where the impact of SDT is more prominent, the good subthreshold characteristics of the [100] direction tends to yield the best performance.

Fig. 9 compares the impact of drain doping on $I_{ON}$ for the [100] cases. Similar trends are observed for the other orientations. To limit SDT degradation, lower optimum drain doping ($N_D = 1\text{-}2 \times 10^{18}$ cm$^{-3}$, $2\text{-}3 \times 10^{18}$ cm$^{-3}$) is observed at $L = 12$ nm compared to $L = 20$ nm ($N_D = 2\text{-}4 \times 10^{18}$ cm$^{-3}$, $7\text{-}8 \times 10^{18}$ cm$^{-3}$) for n- and pTFET, respectively. Lower drain doping tends to lead to non-degeneracy at the drain side (SI Fig. 12a) which limits the tunneling window in the on-state (SI Fig. 12c). When the channel barrier is lowered below the drain CB, further increasing $V_G$ does not increase the available tunneling windows and the current saturate (nFET case, pTFET is similar applying the usual symmetry). The current often decreases, leading to a negative differential resistance (NDR) region (SI Fig. 12b), as a more positive $V_G$ further depletes the source, increasing the tunneling distance and decreasing $T_{ON}$. The maximum tunneling windows to contribute to the current is the energy region between the source and drain Fermi levels and is equal to the applied drain to source potential $V_{DS}$ (in eV) which is typically equal to the supply voltage $V_{DD}$ in a digital circuit. If the drain band edge is not degenerated however the maximum tunneling window will be smaller. This results in an early onset of current saturation in the $I_D(V_G)$ characteristics at low $V_D$ and degrades the output characteristics. At same $N_D$, this effect is more detrimental for the pTFET due to the higher III-V VB DOS with regards to the conduction band (CB).

The impact of the NW cross-section is shown in Fig. 7. For $L = 20$ nm, a diameter of 5.5 nm is optimal with regards to electrostatic control, confinement, and area for current flow both for n- and pTFETs. At $L = 12$ nm, the improved electrostatic control and increased EM of the $d = 4.3$ nm NW help suppressing SDT and work best for the nTFET. For the pTFET cases, despite the better SS, the higher DOS at the drain side combined with low optimum $N_D$ results in an earlier onset of $I_{ON}$ saturation effects and degraded $I_{ON}$ at $V_{DD} = 0.3$V, compared to the $d = 5.5$ nm cases.

To reduce the impact of SDT at $L = 12$ nm, we also assess the possibility of a different material choice (Fig. 10). 2 different configurations using InGaAs are investigated for the nTFET case. The first (InGaAs-Ch) replaces the InAs channel and drain by In$_x$Ga$_{1-x}$As. A 4.2 nm long InAs pocket is kept at the source channel-junction to boost $T_{ON}$ (SI Fig. 13). The second configuration (InGaAs-BL) keeps the InAs channel and drain but uses an InGaAs barrier layer at the drain side (Fig. 10b). Due to the larger InGaAs EM and BG, both configurations are effective to reduce SDT, but also suffer from $T_{ON}$ and current degradation at larger $V_G$. Owing to the effectiveness of the additional barrier to suppress SDT, we found that the InGaAs-BL case using an indium rich InGaAs (x ~ 90 %) gives the best trade-off and can boost performance for very low voltage applications (up to $V_G$ of about 0.25 V in Fig. 10a). Finally, Fig. 11, and Table I benchmark the 12 and 20 nm TFET GAA designs vs. a GAA Si nMOSFET at $I_{OFF} = 50$ pA/μm in terms of $I_{ON}$ and energy delay, highlighting the superior LP performance of the 20 nm GAA III-V TFET design both for $V_{DD} = 0.3$ and 0.5 V.

## DISCUSSION

In this work, we report the capability to simulate in a quantum mechanical tight-binding atomistic fashion NW devices featuring several hundred to millions of atoms and diameter up to 18 nm. Such simulations go far beyond what is typically affordable with today's supercomputers using a traditional real space TB Hamiltonian technique. We have employed an innovative TB mode space technique instead and demonstrate large speedup (up to 10,000×) while keeping good accuracy (error < 1%) compared to the RS NEGF method. Such technique and capability open new avenues to explore and understand the physics of nanoscale and mesoscopic devices dominated by quantum effects.

In particular, we have applied our MS method to the technological relevant case of band-to-band tunneling in III-V nanowire MOSFETs and broken gap heterojunction TFETs. We demonstrate an accurate match of simulated BTBT currents to experimental measurements in a [111] InAs NW having a 12 nm diameter and a 300 nm long channel. We apply the predictivity of our TB MS simulations and report the first in-depth atomistic study of the scaling potential of III-V GAA nanowire heterojunction TFETs quantifying the benefits of this technology for low-power, low-voltage CMOS application. It is shown that both n- and pTFET performances are best above 20 nm gate length for a diameter of 5.5 nm which features the best trade-off between electrostatic control and confinement effects. Such a configuration could benefit from up to 50% gain in the [111] crystal orientation at $V_{DD} = 0.3$ V and $I_{OFF} = 50$ pA/μm. In a low power ITRS 2.0 horizontal GAA beyond 5 nm node design rule however, where the gate length is restricted to 12 nm,[2] a [100] orientation is best, but features 2.3 to 3× $I_{ON}$ degradation and 1.9 to 2.4× energy-delay product (ETP) degradation compared to the 20 nm GAA design. The 20 nm GAA TFET design features significant $I_{ON}$ and ETP performance gain over a Si NW MOSFET of 58× and 56× respectively.

## METHODS

### Tight Binding Mode Space Model

To switch from the original real space of size N to the reduced mode space of size n, a unitary transformation basis composed of the n chosen orthonormal basis eigenvectors { $\Psi_i$ } in the N-dimensional basis needs to be constructed. In matrix notation, any approximate MS quantity, e.g. the Hamiltonian of the NW slab is expressed as:

$h = U^\dagger H U$, with $U = ( \Psi_1, \Psi_2, …, \Psi_n )$ (1)

$U$ is the transformation matrix of size $N \times n$, $H$ is the real space Hamiltonian matrix of size $N \times N$ and $h$ is the related $n \times n$ MS matrix. In the text we will use capital letters for RS quantities, while MS quantities will be represented by small letters.

*1) Initial Sampling:*

Assuming we have selected a number of representative eigenvectors, the real and imaginary part of the Bloch states are taken as $n_0$ columns of a $N \times n_0$ real matrix $U_0$ to create a real-valued orthogonal basis set[24]. In practice we will orthogonalize the overlap matrix $U_0^T U = c \Lambda c^T$ of the sampled eigenvectors to eliminate any linear combination (i.e. eigenvectors with small eigenvalues $\Lambda_i$) and obtain our initial basis $U_i$ of size $n_i < n_0$:

$$U_i = U_0 c \Lambda^{-1/2} \quad (2)$$

*2) Basis Optimization:*

Assuming a MS basis $\Phi$ of size $n'$ at the $i^{th}$ optimization step, we start with $\Phi = U_i$ for the 1st optimization step. We calculate $\phi$ the additional basis vector as[24]:

$$\phi = \frac{1}{\sqrt{C^T C}} \Xi C \quad (3)$$

where $C$ is an array of $m$ optimized expansion coefficients. For the $N \times m$ trial basis matrix $\Xi$, following, [24] we choose to orthogonalize the columns of the $N \times 2n'$ matrix:

$$[(1 - \Phi\Phi^T)H_0\Phi, (1 - \Phi\Phi^T)(W + W^T)\Phi]. \quad (4)$$

where $H_0$ is the tight-binding Hamiltonian of the isolated NW slab and $W$ contains the coupling terms to the atomic orbitals of the next unit structure. Determining $C$ is done by minimizing the following real variational functional: [24]

$$\Delta F(C) = \frac{1}{2n_z} \sum_{i=1}^{n_q} \sum_{k=1}^{2n_z} \frac{C^T A(q_i, z_k) C}{C^T B(q_i, z_k) C} (z_k - \varepsilon_c)(C^T C - 1)^2 \quad (5)$$

$$A(q, z) = 1_{m \times m} + \Xi^T H(q) \Phi [z - h(q)]^{-2} \Phi^T H(q) \Xi \quad (6)$$

$$B(q, z) = z 1_{m \times m} - \Xi^T H(q) \Xi$$
$$- \Xi^T H(q) \Phi [z - h(q)]^{-1} \Phi^T H(q) \Xi \quad (7)$$

where $\varepsilon_C = (E_H + E_L)/2$ and $\rho = (E_H - E_L)/2$ are the center and radius of the contour $C$ in the complex $z$ plane on which the $2n_z$ points $z_k = \varepsilon_C + \rho e^{\frac{i\pi}{n_z}(k - \frac{1}{2})}$ are chosen.

$\Delta F$ measures the changes in the unphysical part of the band structure at $n_q$ chosen wave numbers $q_i$ brought about by the addition of 1 particular basis vector $\phi$.

### Experimental InAs NW fabrication process:

Unintentionally doped InAs NWs with diameters in the range of 7-15 nm were grown by MOCVD using the VLS technique. A hybrid fabrication flow combining growth, high-k, and metal gate formation in a vertical flow and completion of device fabrication in a lateral flow was adopted to enable simple processing, while demonstrating the potential for InAs NW technology. SI Fig. 5 summarizes the process flow. After growth, a $ZrO_2$ high-k with $EOT \sim 0.84$ nm was deposited by ALD, followed by sputtering of W gate metal. NWs were subsequently broken from the growth substrate and transferred to a $SiO_2$/Si chip. The transfer process is random, with the smallest NWs rarely surviving. Devices processed are assumed to have the median diameter as identified by SEM. The median diameter depends on Au particle diameter and growth conditions and is 12 nm for the particular device characteristics shown on Fig. 5b.

E-beam lithography was used to pattern the source/drain (S/D) electrodes. The W gate metal was removed in the S/D contact area by wet etch. This technique allows for self-alignment of the S/D contact to the gate metal and gate-to-S/D distances as small as a few nm. The high-k was removed by dry etch and Ni/Au S/D contacts lifted off. The gate metal pad is similarly patterned by EBL and liftoff.

### NEGF method and speed up consideration:

In the NEGF method, the simulation time is typically dominated by the inversion operations required to calculate the Green's functions G: [20-22]

$$G(E) = (EI - H - \Sigma)^{-1} \quad (8)$$

where E is the scalar energy and $I$ the identity matrix, $H$ the device Hamiltonian, and $\Sigma$ the self-energy that include the interaction terms with the semi-infinite leads and the possible scattering interactions terms (e.g. with phonons) are matrices of rank $N_{tot}$, the total number of atoms in the device × the number of orbitals/atoms. The operation scales like $N_{tot}^3$. Typically however, physical systems feature a limited interaction range and the system matrices are sparse. This sparcity is exploited in the recursive Green's function (RGF) algorithm[38,39] to considerably speed up the operation. In this case computing $G$ scales as $N_B^3 \cdot N_X$. $N_B$ being the typical size of the sparse blocks while $N_X$ being the number of sparse block along the channel direction of the wire.

Assuming a nearest neighbor TB and that the self-energy terms do not include any non-local components (e.g. ballistic or local scattering only) (which is less favorable case for MS/RS due to a sparsity loss in MS as we shall see below), the Hamiltonian in RS is sparse down to a single atomic layer. After transforming the full device Hamiltonian to MS using for each slab the computed slab eigenvector transformation matrices, the Hamiltonian sparcity is only preserved down to the slab level, e.g. 4 atomic planes in a [100] diamond or zincblende crystal structure. Any further sparcity is lost during the transformation and the MS transformed $H$ is typically a full dense matrix. The theoretical maximum speed up, $s_{max}$, is therefore expressed as:

$$s_{max} = r^{-3}/sl^2 \quad r = n_{tot}/N_{tot} \quad sl = N_x/n_x \quad (9)$$

where we defined the reduction ratio $r$, that expressed the size reduction between the MS and RS model Hamiltonian and the sparcity loss factor $sl$, that expressed the possible loss of sparcity of the MS H models vs. the RS one. In our case $sl = 4$.

**Data availability:** The data that support the findings of this study are available from TSMC but restrictions apply to the availability of these data, which were used under license for the current study, and so are not publicly available. Data are however available from the authors upon reasonable request and with permission of TSMC.

REFERENCES

1. Moore G. E. Cramming More Components onto Integrated Circuits. Electronics, 114–117 (1965).
2. ITRS 2015.2.0: http://www.semiconductors.org/news/2016/07/08/press_releases_2015/international_technology_roadmap_for_semiconductors_examines_next_15_years_of_chip_innovation/
3. Colinge J.-P., Lee C.-W., Afzalian A., Dehdashti Akhavan N., Ran Y., Ferain I., Razavi P., O'Neill B., Blake A., White M., Kelleher A.-M., McCarthy B. & Murphy R. Nanowire transistors without junctions. Nature Nanotechnology **5**, 225-229 (2010).
4. van Dal M. J. H., Vellianitis G., Doornbos G., Duriez B., Shen T. M., Wu C.C., Oxland R., Bhuwalka K., Holland M., Lee T. L., Wann C., Hsieh C. H., Lee B. H., Yin K. M., Wu Z. Q., Passlack M. & C. H. Diaz Demonstration of scaled Ge p-channel FinFETs integrated on Si. Int. Electron Devices Meeting (IEDM), 521-524 (2012).
5. Kim D.-H., Kim T.-W., Baek RH., Kirsch P. D., Maszara W., del Alamo J. A., Antoniadis D. A., Urteaga M., Brar B., Kwon HM., Shin C.-S., Park W.-K., Cho Y.-D., Shin SH., Ko DH. & Seo K.-S. High performance III-V devices for future logic applications. Int. Electron Devices Meeting (IEDM), 578-581 (2014).
6. Novoselov K. S., Jiang D., Schedin F., Booth T. J., Khotkevich V. V., Morozov S. V. & Geim A. K. Two-dimensional atomic crystals PNAS **102**, 10451–10453 (2005).
7. Sakurai T. Perspective of low power VLSI's. IEICE Trans. Electron **E87-C**, 429-436 (IEICE, 2004).
8. Ionescu A. M. & Riel H. Tunnel field-effect transistors as energy-efficient electronics switches. Nature **479**, 329-337 (2011).
9. Abelé N., Fritschi R., Boucart K., Casset F., Ancey P. & Ionescu A. M. Suspended-Gate MOSFET: bringing new MEMS functionality into solid-state MOS transistor. Int. Electron Devices Meeting (IEDM), 479-481, 2005.
10. Salahuddin S. & Datta S. Use of Negative Capacitance to Provide Voltage Amplification for Ultralow Power Nanoscale Devices. Nanoletters **8**, 405 (2008).
11. DasGupta S., Rajashekhar A., Majumdar K., Agrawal N., Razavieh A., S. Trolier-McKinstry & Datta S. Sub-kT/q Switching in Strong Inversion in PbZr0.52Ti0.48O3 Gated Negative Capacitance FETs. IEEE Journal on Exploratory Solid-State Computational Devices and Circuits 1, 43-48 (2015).
12. Gopalakrishnan K., Griffin P. B. & Plummer J. D., I-MOS: A novel semiconductor device with subthreshold slope lower than kT/q. Int. Electron Devices Meeting (IEDM), 289–292 (2002).
13. Afzalian A., Colinge J.P. & Flandre D. Physics of Gate Modulated Resonant Tunneling (RT)-FETs: Multi-Barrier MOSFET for Steep Slope and High On-Current. Solid-State-Electron. **59**, 50-61 (2011).
14. Gnani E., Reggiani S., Gnudi A., & Baccarani G. Steep-slope nanowire FET with a superlattice in the source extension. Solid-State-Electron. **65-66**, 108-113 (2011).
15. Appenzeller J., Lin Y.-M., Knoch J. & Avouris Ph. Band-to-band tunneling in carbon nanotube field-effect transistors. Phys. Rev. Lett. **93**, 196805 (2004).
16. Memisevic E., Svensson J., Hellenbrand M., Lind E. & Wernersson L.-E. Vertical InAs/GaAsSb/GaSb Tunneling Field-Effect Transistor on Si with S = 48 mV/decade and Ion = 10 µA/µm for Ioff = 1 nA/µm at Vds = 0.3 V. Int. Electron Devices Meeting (IEDM), 19.1.1-19.1.4 (2016).
17. Avci U.E., Morris D. H., Hasan S., Kotlyar R., Kim R., Rios R., Nikonov D.E. & Young I. A., Energy Efficiency Comparison of Nanowire Heterojunction TFET and Si MOSFET at Lg=13nm, Including P-TFET and Variation Considerations. Int. Electron Devices Meeting (IEDM), 33.4.1-33.4.4 (2013).

18. Afzalian A., Passlack M. & Yeo Y.-C. Scaling perspective for III-V broken gap nanowire TFETs: An atomistic study using a fast tight-binding mode-space NEGF model. Int. Electron Devices Meeting (IEDM), 30.1.1-30.1.4 (2016).
19. Gummel H.K. A self-consistent iterative scheme for one-dimensional steady state transistor calculations. IEEE Trans. Elect. Dev. **11**, 455 - 465 (1964).
20. P. Keldysh Diagram technique for nonequilibrium processes. Sov. Phys. JETP 20, 1018 (1965).
21. P. Kadanoff and G. Baym, Quantum Statistical Mechanics, Benjamin, New York, 1962.
22. S. Datta Nanoscale device modeling: the Green's function method. Superlattices and Microstructures, **28**, 253-278 (2000).
23. Slater J. C. & Koster G. F. Simplified LCAO Method for the Periodic Potential Problem. Phys. Rev. **94**, p. 1498 (1954).
24. Mil'nikov G., Mori N. & Kamakura Y., Equivalent transport models in atomistic quantum wires. Phys. Rev. B **85**, 035317 (2012).
25. Steiger S., Povolotskyi M., Park H.-H., Kubis T. & Klimeck G., NEMO5: A Parallel Multiscale Nanoelectronics Modeling Tool. IEEE TNANO **10**, 1464 (2011).
26. Luisier M. & Klimeck G., Atomistic full-band design study of InAs band-to-band tunneling field-effect transistors. IEEE Elec. Dev. Letters **30**, 602 (2009).
27. A. Afzalian, Passlack M & Yeo Y.-C Atomistic simulation of gate-all-around GaSb/InAs nanowire TFETs using a fast full-band mode-space NEGF model. International Symposium on VLSI Technology, Systems and Application (VLSI-TSA), 1-2 (2016).
28. Kayer M.A. & Lake R.K. Effects of heavily doped source on the subthreshold characteristics of nanowire tunneling transistors DRC, 51 (2011).
29. Koswatta S.O., Koester S.J. & Haensch W. On the Possibility of Obtaining MOSFET-Like Performance and Sub-60-mV/dec Swing in 1-D Broken-Gap Tunnel Transistors. IEEE TED **57**, 3222 (2009).
30. Venugopal R., Ren Z., Datta S., Lundstrom M. S. & Jovanovic D. Simulating quantum transport in nanoscale transistors: Real versus mode-space approaches. J. Appl. Phys. **92**, 3730–3739 (2002).
31. Wang J., Polizzi E. & Lundstrom M., A three-dimensional quantum simulation of silicon nanowire transistors with the effective-mass approximation. J. Appl. Phys. **96**, 2192-2203 (2004).
32. Luisier M., Schenk A. & Fichtner W. Quantum transport in two-and three-dimensional nanoscale transistors: coupled mode effects in the nonequilibrium Green's function formalism. J. Appl. Phys. **100**, 043713 (2006).
33. Jin S., Park Y.J. & Min H. S. A three-dimensional simulation of quantum transport in silicon nanowire transistor in the presence of electron-phonon interactions. J. Appl. Phys. **99**, 123719 (2006).
34. Afzalian A., Dehdashti Akhavan N., Lee C.-W., Ran Y., Ferain I., Razavi P & Colinge J.-P. A new F(ast)-CMS NEGF Algorithm for efficient 3D simulations of Switching Characteristics enhancement in constricted Tunnel Barrier Silicon Nanowire MuGFETs. J. of Comp. Electronics **8**, 287-306 (2009).
35. Afzalian A. Computationally Efficient self-consistent Born approximation treatments of phonon scattering for Coupled-Mode Space Non-Equilibrium Green's Functions. J. Appl. Phys. **110**, 094517 (2011).
36. Afzalian A., Huang J., Ilatikhameneh H., Charles J., Lemus D., Bermeo Lopez J., Perez Rubiano S., Kubis T., Povolotskyi M., Klimeck G., Passlack M. & Yeo Y.-C. Mode space tight binding model for ultra-fast simulations of III-V nanowire MOSFETs and heterojunction TFETs, International Workshop of Computational Electronics (IWCE), 1-3 (2015).
37. Vasen T., Ramvall P., Afzalian A., Thelander C., Dick K.A., Holland M., Doornbos G., Wang S.W., Oxland R., Vellianitis G., van Dal M.J.H., Duriez B., Ramirez J.-R., Droopad R., Wernersson L.-E., Samuelson L., Chen T.-K., Yeo Y.-C. & Passlack M. InAs Nanowire GAA n-MOSFETs with 12-15 nm Diameter. Symposia on VLSI Technology and Circuits (VLSI) (2016).
38. Lake, R., Klimeck, G., Bowen, R., Jovanovic, D. Single and multiband modeling of quantum electron transport through layered semiconductor devices. J. Appl. Phys. **81**, 7845–7869 (1997).
39. Svizhenko, A., Anantram, M., Govindan, T., Biegel, R., Venugopal, R. Two-dimensional quantum mechanical modeling o.f nanotransistors. J. Appl. Phys. 91, 2343–2354 (2002).
40. Nagavarapu, Jhaveri R. & Woo J. C. S The Tunnel Source (PNPN) n-MOSFET: A Novel High Performance Transistor. IEEE Trans. Elect. Dev. **55**, 1013-1019 (2008).

41. Verreck D., Verhulst A. S., Sorée B., Collaert N., Mocuta A., Thean A. & Groeseneken G. Improved source design for p-type tunnel field-effect transistors: Towards truly complementary logic APL **105**, 243506 (2014).

**ACKNOWLEDGMENTS**

The authors thank Dr. Y. C. Sun for management support, and Dr. J. Huang for its contribution to the code development

**Author contributions**
A.A. worked on the theory, including code and algorithm development and optimization, MS bases derivation, performed the simulations, took part in the experimental project definition and wrote the paper. T.V. performed the experimental device processing and the electrical characterization. P. R. was involved in the MOCVD growth of the NW and the physical characterization. M.P. was involved in the experimental project definition and direction, data analysis and improved the manuscript text.

**Competing financial interests:**
The authors declare no competing financial interests.

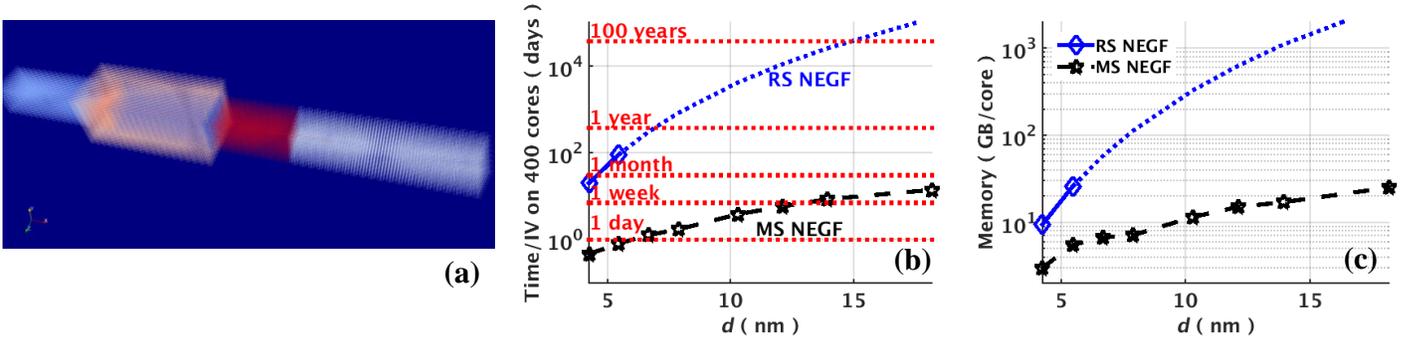

**Fig. 1.** (a) Atomistic view of a [111] GAA NW nTFET with a diameter of 5.5 nm and featuring several 100,000 atoms. (b) Simulation time per IV and (c) peak memory per core (typical values) to simulate an InAs/GaSb NW TFET vs. NW diameter using a real space vs. mode space NEGF and a $sp^3s^*SO$ (SO = spin-orbit coupling, 10 orbitals/atom) TB basis Hamiltonian on a 400 cores cluster. RS NEGF values are measured for $d = 4$ and 5.5 nm and extrapolated using a $d^6$ (time) and $d^4$ (memory) law for larger diameter (see section results).

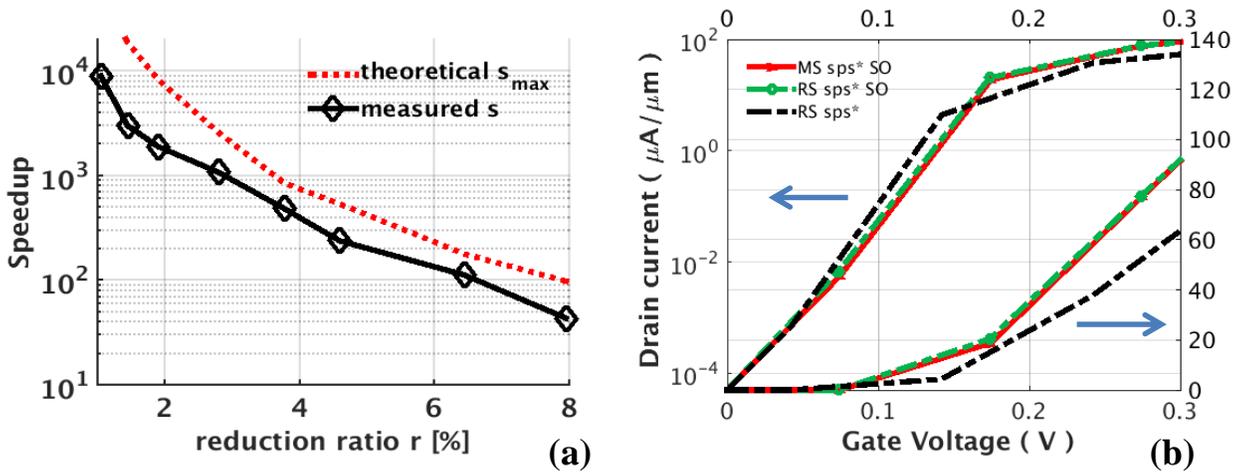

**Fig. 2:** (a) MS to RS speed up factor vs. reduction ratio extracted from Fig. 1.b. Maximum theoretical speed up from eq. (9) and speed up achieved in our implementation. Using the MS models, reduction ratio ranging from 8% (at $d = 4$ nm) down to 1% (at $d = 18$ nm) were observed and speed up factors $s$ ranging from 40 to 10,000 were achieved, when compared to the initial TB models. (b) $I_D(V_G)$ characteristics of a 15 nm long InAs-GaSb PNIN n-TFET computed from the original (RS) TB model and from optimized InAs and GaSb MS basis (speed up > 150×, error < 1%). The $I_D(V_G)$ from a RS model neglecting SO (speed up 8×, error > 50%) is also shown. The current is normalized by the NW perimeter. $V_D = 0.3V$.

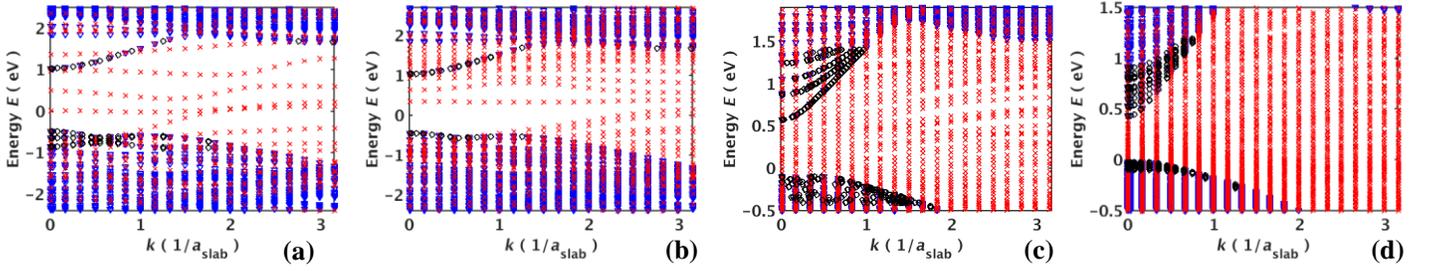

**Fig. 3.** *E-k* dispersion of [100] InAs NW slabs computed from the original (Real Space) TB model (▼) and from MS bases (×) for different NW diameters *d* and TB bases. The initial (uncleaned) MS bases are used for (a) a *d* = 1.8 nm $sp^3s^*$ basis (5 orbitals/atom), (b) a *d* = 1.8 nm $sp^3s^*SO$ basis (10 orbitals/atom), (c) a *d* = 5.5 nm $sp^3s^*SO$ basis and (c) a *d* = 18 nm $sp^3s^*SO$ basis. The *E-k* points used to construct the initial MS bases (from the related eigenvectors) are indicated by black circles. The number of unphysical modes (UM) increases significantly as diameter and number of orbitals of the TB basis increase.

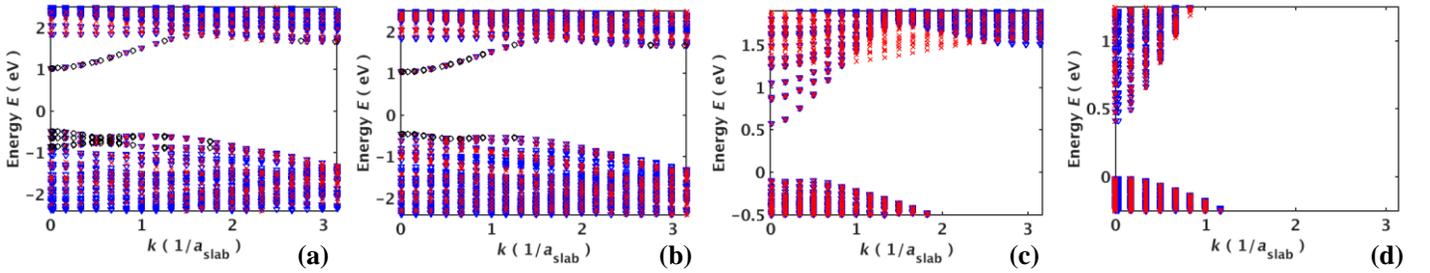

**Fig. 4.** *E-k* dispersion of [100] InAs NW slabs computed from the original (Real Space) TB model (▼) and from MS bases (×) for different NW diameters *d* and TB bases. The optimized (cleaned) MS bases are used for (a) a *d* = 1.8 nm $sp^3s^*$ basis (5 orbitals/atom) after 24 optimization steps, (b) a *d* = 1.8 nm $sp^3s^*SO$ basis (10 orbitals/atom) after 60 optimization steps, (c) a *d* = 5.5 nm $sp^3s^*SO$ basis after 75 optimization steps and (d) a *d* = 18 nm $sp^3s^*SO$ basis after 480 optimization steps.

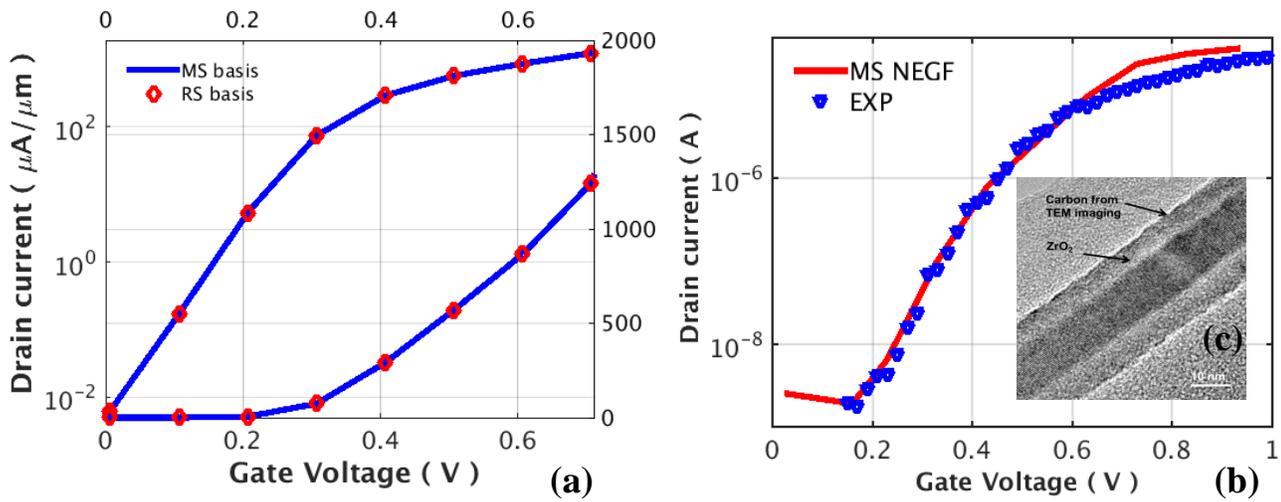

**Fig. 5:** *(a)* $I_D(V_G)$ characteristics of a [100] $d$ = 5.5 nm InAs n-MOSFET computed from the original (RS) TB model and from an optimized InAs MS basis (speed up 150×). $L$ = 15 nm. **(b)** $I_D(V_G)$ characteristics of a [111] $d$ =12 nm InAs nMOS transistor measured and simulated using a ballistic TBMS – NEGF model from an optimized $sp^3s*\_SO$ InAs MS basis. $V_D$ = 0.5V. $L$ = 300 nm. **(c)** High-resolution TEM image along a NW with about 4 nm $ZrO_2$ high-k and $d$ = 12 nm.

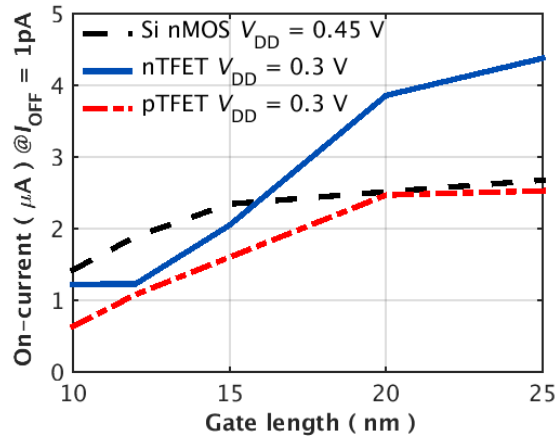

**Fig. 6.** Impact of gate length on $I_{ON}$ (current /wire) of optimized InAs/GaSb GAA NW n- and pTFETs and Si GAA NW nMOSFETs. Gate oxide: 1.8 nm $Al_2O_3$ oxide. $I_{OFF}$ = 1 pA/ wire, $V_{DD}$ = 0.3 V for the TFETs and 0.45 V for the MOSFET.

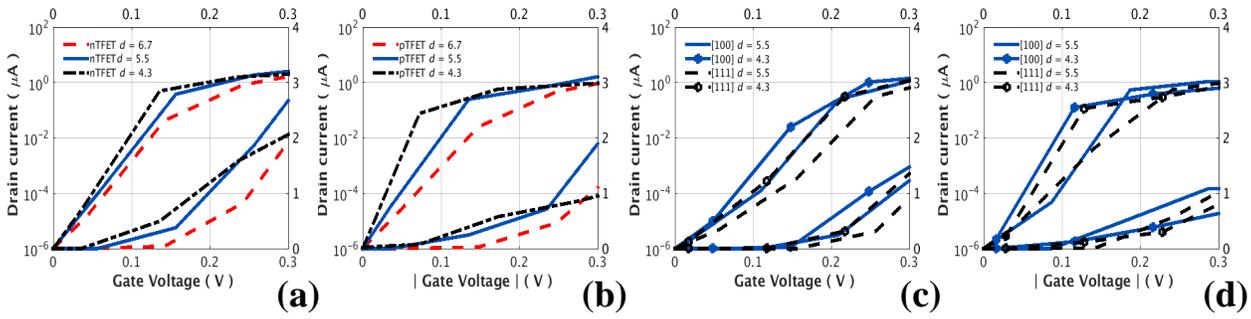

**Fig. 7.** Impact of diameter on $I_D(V_G)$ characteristics of optimized InAs/GaSb GAA NW TFETs for (a) $L$ = 20 nm [100] nTFETs, (b) $L$ = 20 nm [100] pTFETs. $V_D$ =0.3V. The gate work function of the individual devices were adjusted to have $I_{OFF}$ = 1pA at 0V.

**Fig. 20.** Impact of diameter on $I_D(V_G)$ characteristics of optimized InAs/GaSb GAA NW TFETs for (a) $L$ = 12 nm [100] and [111] nTFETs, and (b) $L$ = 12 nm [100] and [111] pTFETs. $V_D$ =0.3V. $I_{OFF}$ = 1pA.

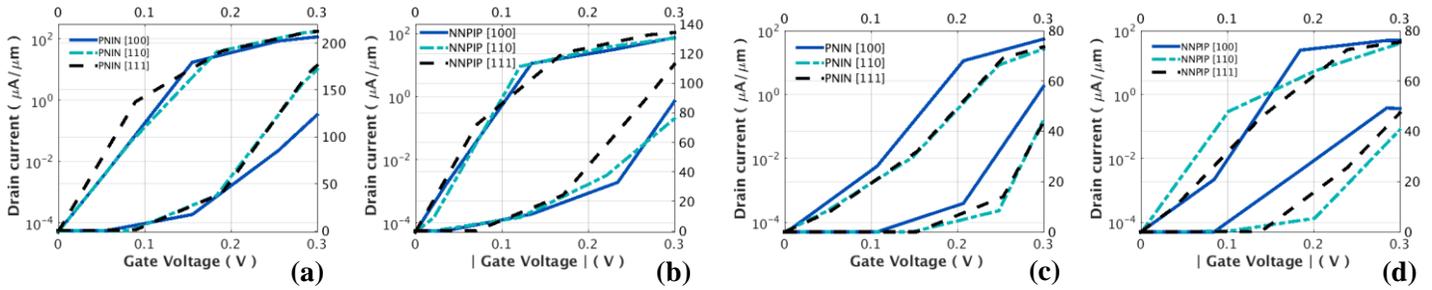

**Fig. 8.** Impact of channel orientation on $I_D(V_G)$ characteristics of optimized InAs/GaSb GAA NW TFETs for (a) $L$ = 20 nm nTFETs, (b) $L$ = 20 nm pTFETs, (c) $L$ = 12 nm nTFETs, (d) $L$ = 12 nm pTFETs.

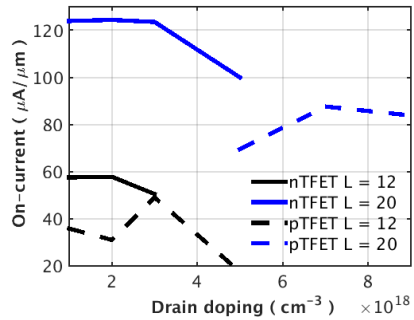

**Fig. 9.** $I_{ON}$ vs. drain doping for [100] InAs/GaSb $L$ = 12 and $L$ = 20 nm GAA NW TFETs. $V_{DD}$ = 0.3 V. $I_{OFF}$ = 50 pA/μm.

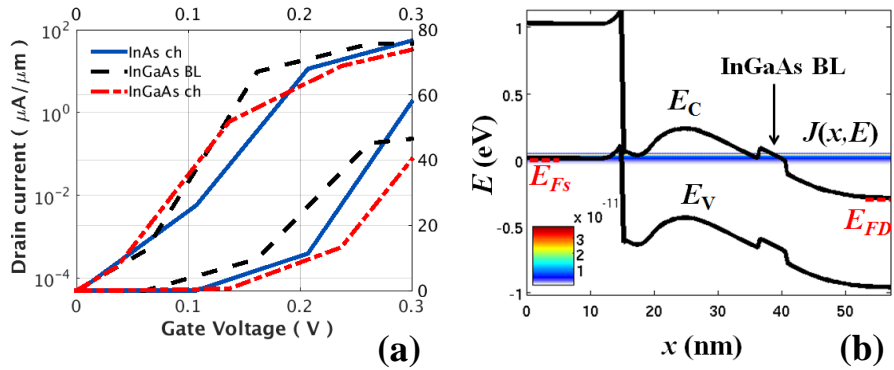

Fig. 10. (a) Impact of InGaAs channel or InGaAs barrier layer at drain side on $I_D(V_G)$ characteristics of optimized LP III-V $L$ = 12 nm GAA NW n HTFETs. The reference case (InAs channel) is also shown. (b) Current spectrum $J(x, E)$ in A/eV (surface plot) and band edges (solid lines) along the transport direction $x$ for the $L$ = 12 nm, InGaAs-BL PNIN TFET of Fig. 10a at $V_G$ = 50 mV. S and D Fermi level positions are also indicated (dashed lines).

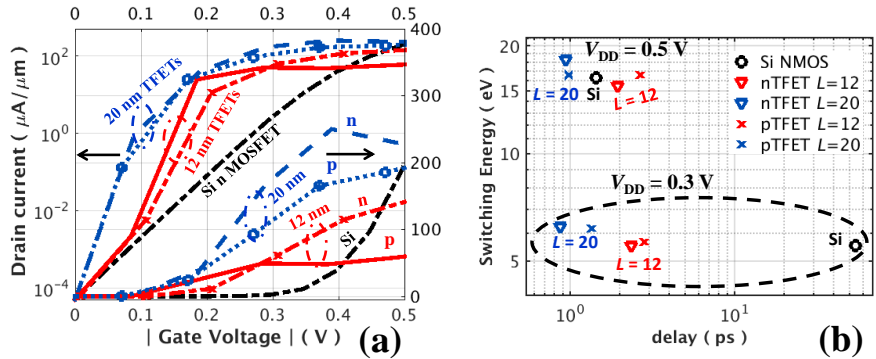

Fig. 11. (a) $I_D(V_G)$ characteristics of optimized Si NW GAA nMOSFETs, and InAs/GaSb $L$ = 12 and $L$ = 20 nm GAA NW n- and pTFETs. $V_{DD}$ = 0.3 V. $I_{OFF}$ = 50 pA/μm. (b) Switching Energy *vs.* delay (unloaded inverter cell with 3 NW/device) at $V_{DD}$ = 0.3 and 0.5 V, for optimized InAs/GaSb $L$ = 12 and $L$ = 20 nm GAA NW n- and pTFETs and optimized Si NW GAA nMOSFETs at $I_{OFF}$ = 50 pA/μm.

|  | 12 nm GAA III-V NW TFET | 20nm GAA III-V NW TFET | GAA Si NW nMOSFET |
|---|---|---|---|
| $d$ (nm) | 5.5 | 5.5 | 4.5 |
| $L$ (nm) | 12 | 20 | 12 |
| EOT (nm) | 0.7 | 0.7 | 0.7 |
| Orientation | [100] | [111] | [100] |
| $I_{OFF}$ (pA/µm) | 50 | 50 | 50 |
| $I_{ON}$ [n/p] (µA/µm) $V_{DD}$ = 0.3 V | 58 / 49 | 176 / 114 | 3 / - |
| $I_{ON}$ [n/p] (µA/µm) $V_{DD}$ = 0.5 V | 118 / 90 | 283 / 249 | 201 / - |
| ETP [n/p] ( eV·ps) $V_{DD}$ = 0.3 V | 12.9 / 15.9 | 5.4 / 8.3 | 300 / - |
| ETP [n/p] ( eV·ps) $V_{DD}$ = 0.5 V | 30 / 44 | 17.2 / 16.2 | 23.4 / - |

**Table I. Summary of predicted performance for LP $L$ = 12 and $L$ = 20 nm GAA InAs/GaSb NW TFETs and Si GAA nMOSFET.**

# Supplementary information (SI):

*TB MS NEGF simulation of III-V Broken Gap NW TFETs: Details on the device structure and the use of doping pockets:*

Doping pockets were used to boost the performance of the TFETs.[35,36] SI Fig. 8a and Fig. 1a show the schematic and atomistic details respectively of an optimized GaSb/InAs PNIN GAA NW n-TFET with $d = 5.5$ nm. A $N^+$-pocket is inserted between the GaSb $P^+$-source and the InAs intrinsic channel. SI Fig. 9 demonstrates the benefit of such a $N^+$-pocket by comparing the $I_D(V_G)$ and $SS(V_G)$ characteristics with (PNIN) and without (PIN) the pocket at $I_{OFF} = 50$ pA/µm. We used an optimal source doping $N_S = 5 \times 10^{19}$ cm$^{-3}$ for the n-TFET case.

SI Fig. 8b and 8c show the schematic and atomistic details respectively of an optimized GaSb/InAs NNPIP GAA NW p-TFET with $d = 5.5$ nm. In addition to the 1$^{st}$ $P^+$-pocket, which is the equivalent to the $N^+$-pocket in the nTFET case, a second $N^+$-pocket was added in the source of the device. Fig. 10 compares optimized pTFET devices without (NIP) and with 1 (NPIP) or 2 doping pockets. For the pTFET, $N_S$ cannot be as high as for the nTFET. Due to the lower III-V conduction band (CB) DOS with regards to the valence band (VB), an optimal source doping of about $10^{19}$ cm$^{-3}$ has to be used to avoid a strong source degeneracy that is detrimental to SS. Decreasing $N_S$ does however negatively affect the tunneling probability in the on-state, $T_{ON}$. A better trade-off, with performance closed to that of the PNIN nTFET case, can be obtained by further reducing $N_S$ to $5 \times 10^{18}$ cm$^{-3}$ but adding a second highly doped pocket at the source/channel interface to boost $T_{ON}$ similarly to what was observed in a planar device.[36] In an optimized GAA NW device however, owing to the better electrostatic control and the stronger confinement that strongly increases the CB DOS, good performance can already been achieved with only one pocket and SS below 60mV/dec can be obtained without any pocket (SI Fig. 10b), which was not the case for 10 nm thick DG TFETs.[36]

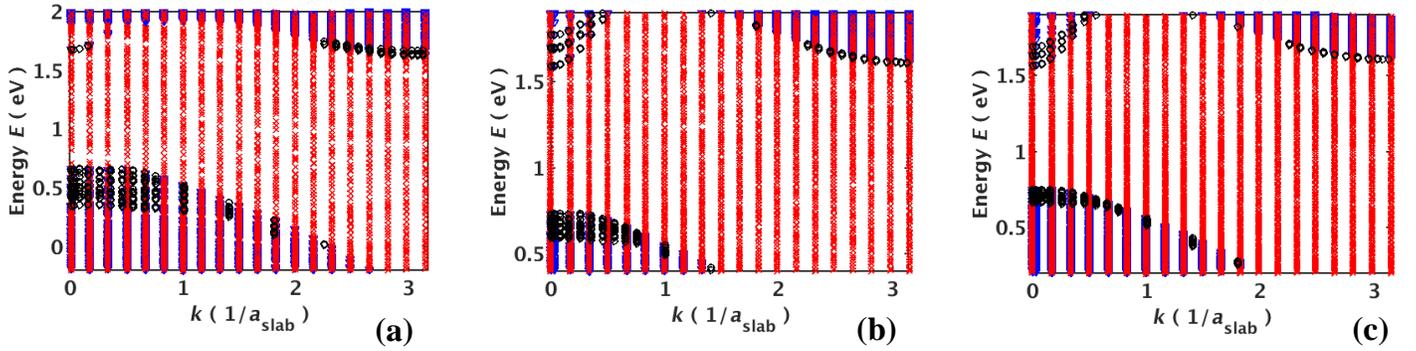

**Fig. 1.** *E-k* **dispersion of [100] GaSb NW slabs computed from the original (Real Space) TB model (▼) and from MS bases (×) for different NW diameters *d*. The initial (uncleaned) MS bases are used for (a) *d* = 5.5 nm, (b) *d* = 10 nm, and (c) *d* = 14 nm. The *E-k* points used to construct the initial MS bases (from the related eigenvectors) are indicated by black circles.** $sp^3s^*SO$ **basis.**

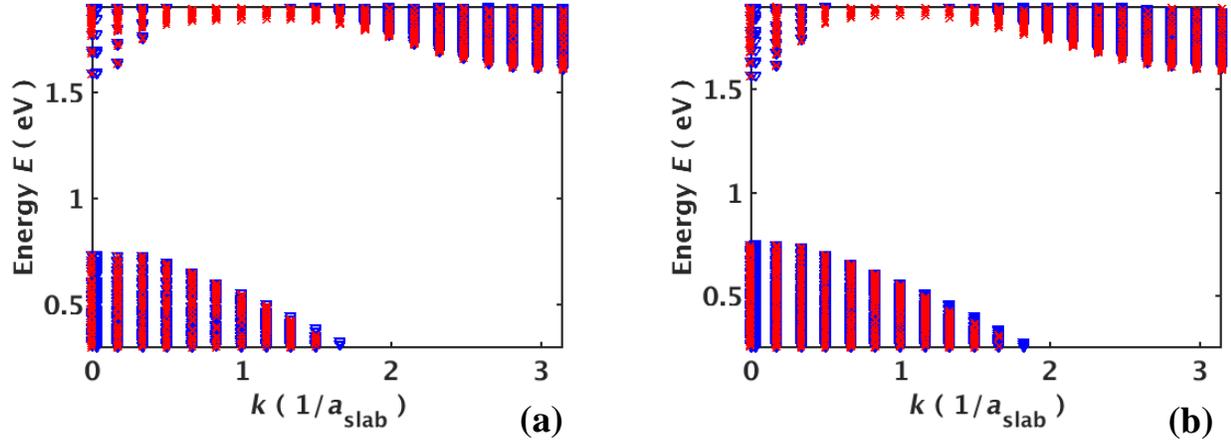

**Fig. 2.** *E-k* dispersion of [100] GaSb NW slabs computed from the original (Real Space) TB model (▼) and from MS bases (×) for different NW diameters *d*. The optimized (cleaned) MS bases are used for (a) a *d* = 10 nm after 290 optimization steps, and (b) a *d* = 14 nm after 370 optimization steps. $sp^3s^*SO$ basis.

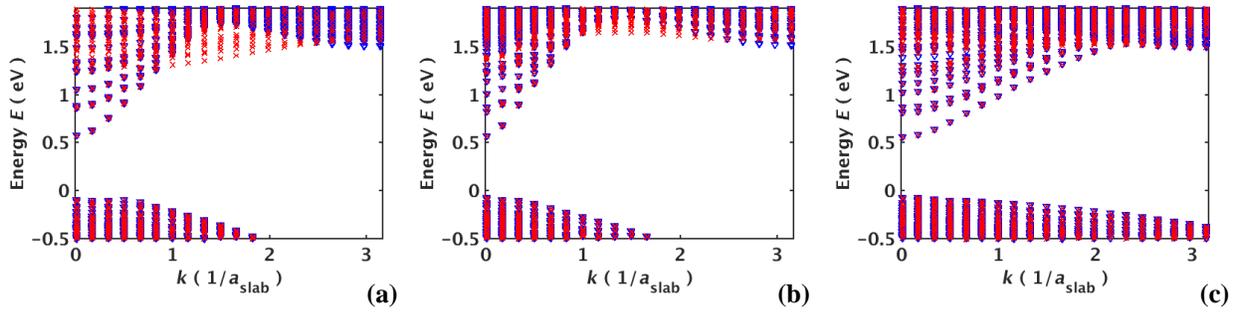

**Fig. 3.** *E-k* dispersion of *d* = 5.5 nm InAs NW slabs computed from the original (Real Space) TB model (▼) and from MS bases (×) for different channel orientations. Optimized (cleaned) MS bases are used for (a) [100], (b) [110], and (c) [111] channel orientation. $sp^3s^*SO$ basis.

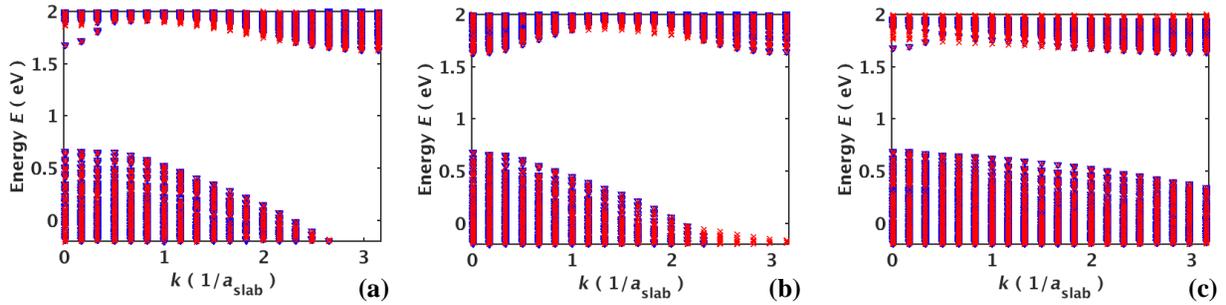

**Fig. 4.** *E-k* dispersion of *d* = 5.5 nm GaSb NW slabs computed from the original (Real Space) TB model (▼) and from MS bases (×) for different channel orientations. Optimized (cleaned) MS bases are used for (a) [100], (b) [110], and (c) [111] channel orientation. $sp^3s^*SO$ basis.

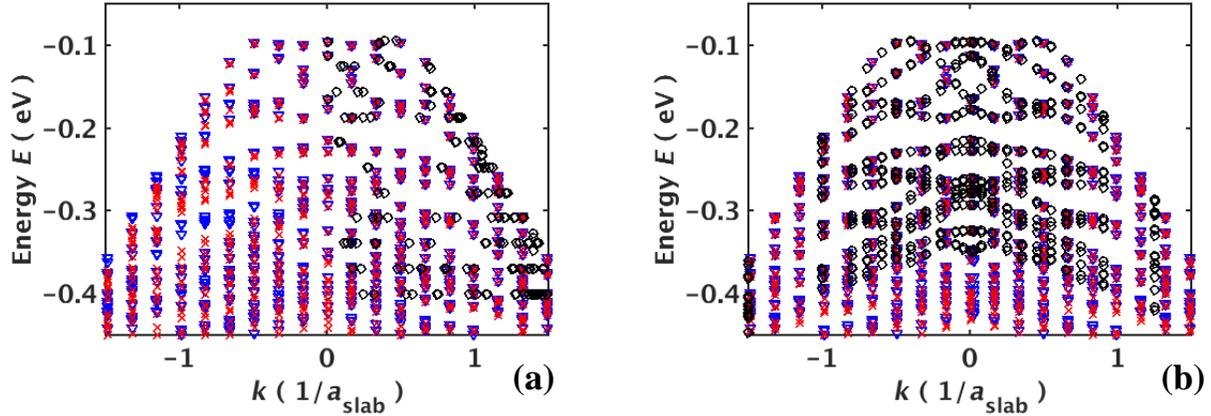

**Fig. 5.** Zoom on the VB *E-k* dispersion of [100] *d* = 4.2 nm $sp^3s^*SO$ basis (10 orbitals/atom) GaSb NW slab computed from the original (Real Space) TB model (▼) and from optimized MS bases (×).The *E-k* points used to construct the initial MS bases (from the related eigenvectors) are indicated by black circles. (a) Only a half of the Brillouin zone (interval $0 - \pi$) is sampled to construct the initial MS basis and a loss of accuracy is observed in the dispersion relation of the optimized MS basis in the other half of the BZ as indicated by the arrow. (b)The full BZ is sampled to construct the MS basis and good accuracy over the full BZ is observed in the optimized MS dispersion relation.

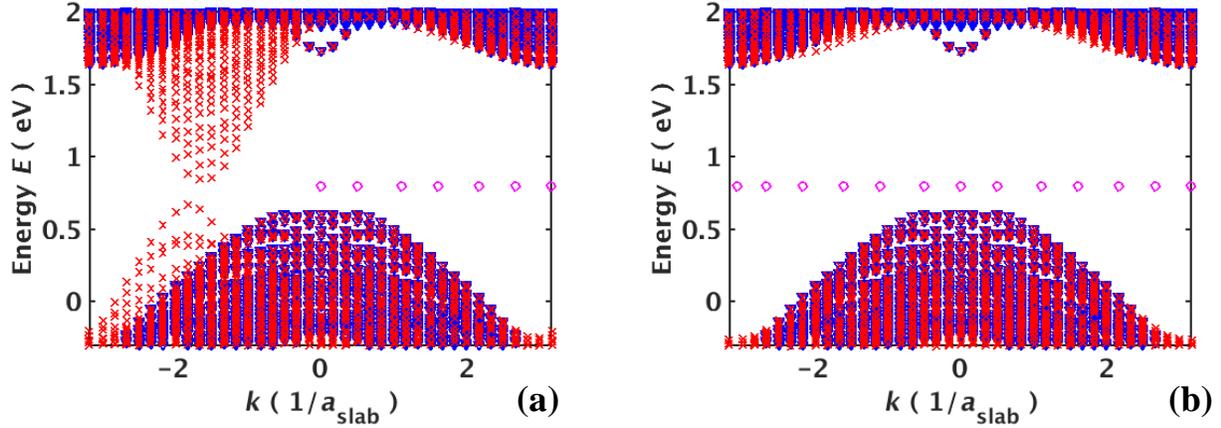

**Fig. 6.** Zoom on the VB *E-k* dispersion of [100] *d* = 4.2 nm $sp^3s^*SO$ bases (10 orbitals/atom) GaSb NW slab computed from the original (Real Space) TB model (▼) and from optimized MS bases (×). The position of the $q_i$ optimization points used to clean the 2 MS bases are indicated by magenta circles (The energy position is arbitrarily chosen). For both bases we have used an initial sampling on the full BZ. (a) Only a half of the Brillouin zone (interval $0 - \pi$) is optimized and some UM remain in the dispersion relation of the optimized MS basis in the other half of the BZ. (b) The full BZ is optimized and the MS basis is cleaned from UM in the chosen energy windows over the full BZ.

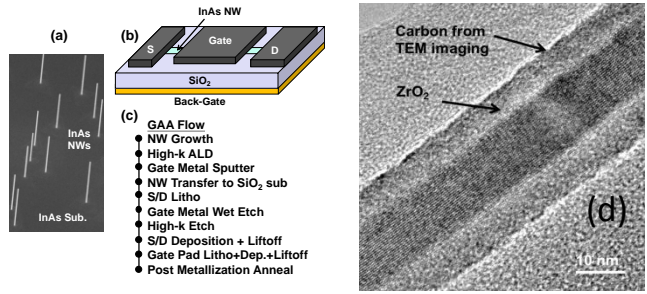

**Fig. 7.** (a) SEM image of as- grown InAs NWs, (b) schematic of completed device structure in lateral flow, and (c) gate all around (GAA) process steps. (d) High-resolution TEM image along a NW with about 4 nm $ZrO_2$ high-k and $d$ = 12 nm.

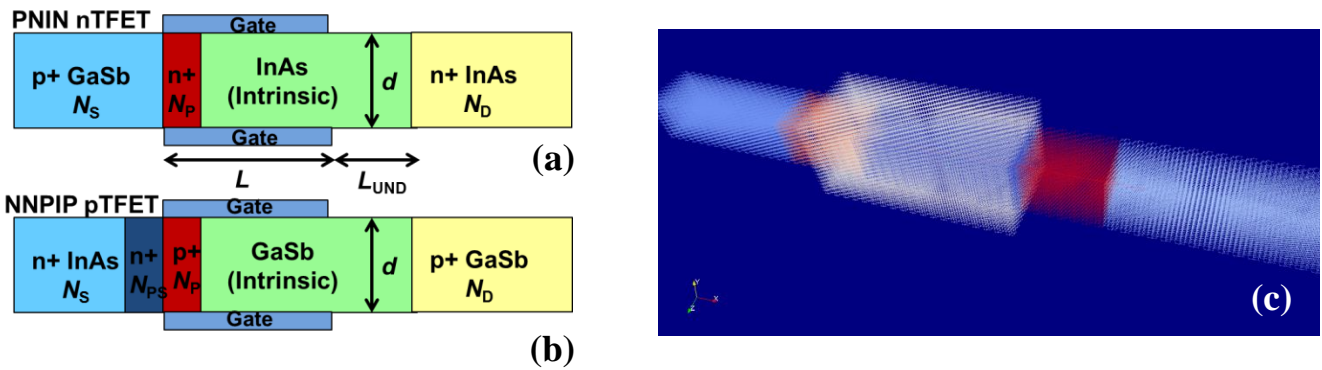

**Fig. 8:** Typical optimized structures of simulated GaSb / InAs NW GAA TFETs. (a) Schematic view of a PNIN nTFET. (b) Schematic view of a NNPIP pTFET. (C) Atomistic view of a [111] pTFET with $d$ = 5.5 nm and featuring several 100,000 atoms. For the atoms, we used a TB $sp^3s^*$ basis including SO coupling.

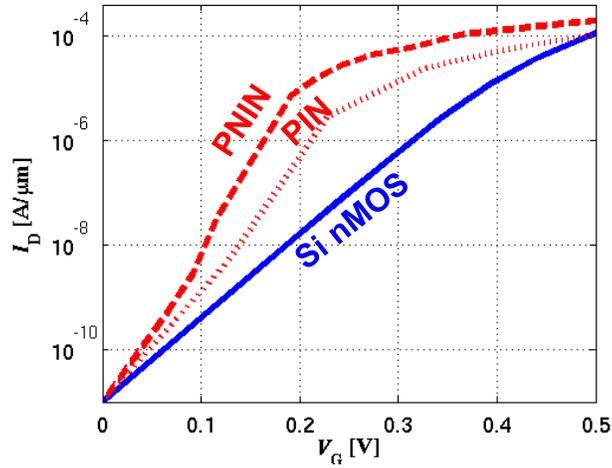

**Fig. 9.** $I_D(V_G)$ characteristics of $d$ = 5.5nm [100] 15 nm long InAs-GaSb GAA nTFET with (PNIN) and without (PIN) a doping pocket. The $I_D(V_G)$ for a L =15nm [100] Si GAA nMOSFET is also shown. $V_D = 0.3V$.

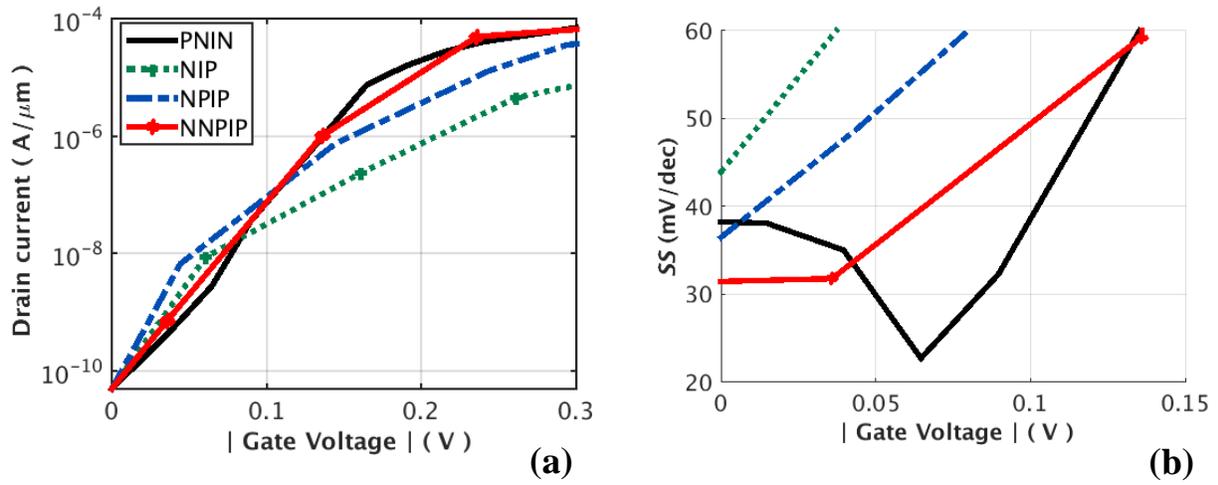

**Fig. 10** (a) $I_D(V_G)$ and (b) $SS(V_G)$ characteristics of $d$ = 5.5nm [100] 15 nm long InAs-GaSb GAA pTFET with 1 (NPIP), 2 (NNPIP) and without (NIP) doping pockets. $V_D = 0.3V$.

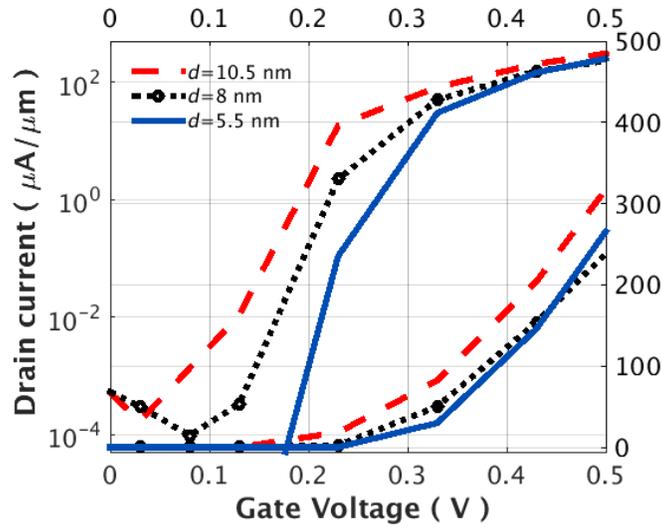

**Fig. 11.** $I_D(V_G)$ characteristics of InAs-GaSb GAA PNIN nTFET for $d = 5.5$, 8 and 10.5 nm. $L = 30$ nm, $N_D = 4 \times 10^{18}$ cm$^{-3}$. $V_D = 0.3$V.

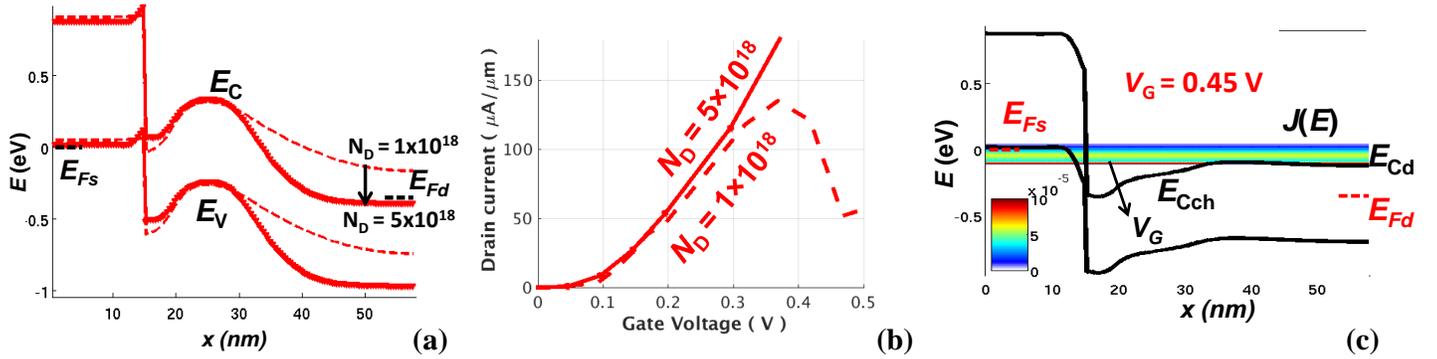

**Fig. 12.** (a) Band edges along the transport direction $x$ in subthreshold regime, and (b) $I_D(V_G)$ characteristics for a $d = 5.5$ nm InAs/GaSb PNIN TFET for $N_D = 1 \times 10^{18}$ cm$^{-3}$ (dashed red lines, non - degenerated drain) and for $N_D = 5 \times 10^{18}$ cm$^{-3}$ (plain red lines, degenerated drain). On-current saturation and NDR region are observed for the non-degenerated case for $V_G$ larger than 0.35V. (c) Current spectrum $J(E, x)$ in A/eV (surface plot) and band edges (-) along the transport direction $x$ for the $N_D = 1 \times 10^{18}$ cm$^{-3}$ PNIN TFET at $V_G = 0.45$ V (in the saturation region). Increasing $V_G$ above 0.35 V did not result in an increase of the available energy windows for tunneling as the current path is limited by the non-degenerated drain CB edge. S and D Fermi level positions are also indicated (dashed lines) in (a) and (c). $L = 15$ nm. $V_D = 0.35$ V. $I_{OFF} = 10$nA/μm.

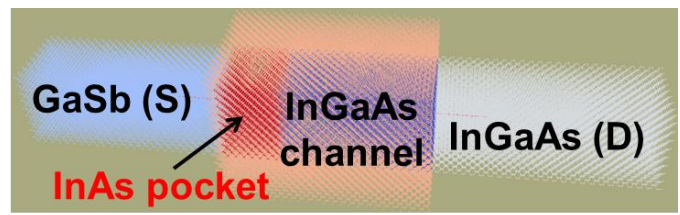

**Fig. 13.** Atomistic structure of the simulated InGaAs channel $L = 12$ nm GAA NW n HTFET.

22